\documentclass[useAMS,usenatbib]{mnras}

\usepackage{graphicx}	
\usepackage{color}	
\usepackage{subcaption}
\usepackage{siunitx}
\usepackage{amsmath}
\usepackage{epstopdf}
\usepackage{bold-extra}

\newcommand{\lbol}{$L_{\textrm{iso}}$}
\newcommand{\Loiii}{$L_{[\textrm{O}\,\textsc{iii}]}$}
\newcommand{\oiii}{${[\textrm{O}\,\textsc{iii}]}$}
\newcommand{\oiiitit}{${[{\rm O}\,\textsc{iii}]}$}
\newcommand{\RFeII}{$R_{\textrm{Fe}\,\textsc{ii}}$} 
\newcommand{\FeII}{Fe~\textsc{ii}}
\newcommand{\Hb}{H$\beta$}
\newcommand{\NeIIIdblt}{[Ne\,{\sc iii}]~$\lambda\lambda$3869, 3967}
\newcommand{\OIIdblt}{[O\,{\sc ii}]~$\lambda\lambda$3726, 3729}
\newcommand{\SIIw}{[S\,{\sc ii}]~$\lambda$4072}

\title[\oiiitit\ bolometric corrections]{Updating quasar bolometric luminosity corrections. III. \hspace{0.5cm}[O {\sc iii}] bolometric corrections}

\author[Pennell et al.]
{\parbox{\textwidth}{Alison Pennell$^{1}$\thanks{E-mail: avp5473@psu.edu, runnoejc@umich.edu},
Jessie C. Runnoe$^{2,3}$\textcolor{blue}{\footnotemark[1]},
M. S. Brotherton$^{4}$}
\vspace{0.4cm}\\ \\
\parbox{\textwidth}{$^{1}$Department of Electrical Engineering, The Pennsylvania State University, 342 Information Sciences and Technology Building, University Park, PA 16802, USA\\
$^{2}$Department of Astronomy \& Astrophysics, and Institute for Gravitation and the Cosmos, The Pennsylvania State University, 525 Davey Lab, University Park, PA 16802, USA\\
$^{3}$Department of Astronomy, University of Michigan, 1085 S. University Ave., Ann Arbor, MI, 48109, USA\\
$^{4}$Department of Physics and Astronomy, University of Wyoming, Laramie, WY 82071, USA
}}

\date{Preprint 2017 January 30}

\pagerange{\pageref{firstpage}--\pageref{lastpage}} \pubyear{2016}

\begin{document}
\label{firstpage}
\pagerange{\pageref{firstpage}--\pageref{lastpage}}
\maketitle

\begin{abstract}
We present quasar bolometric corrections using the $[\textrm{O}\,\textsc{iii}]~\lambda5007$ narrow emission line luminosity based on the detailed spectral energy distributions of 53 bright quasars at low to moderate redshift ($0.0345<z<1.0002$).  We adopted two functional forms to calculate \lbol, the bolometric luminosity determined under the assumption of isotropy: $\textrm{\lbol}=A\,\textrm{\Loiii}$ for comparison with the literature and $\textrm{log(\lbol)}=B+C\,\textrm{log(\Loiii)}$, which better characterizes the data.  We also explored whether ``Eigenvector 1'', which describes the range of quasar spectral properties and quantifies their diversity, introduces scatter into the \Loiii--\lbol\ relationship.  We found that the \oiii\ bolometric correction can be significantly improved by adding a term including the equivalent width ratio \RFeII$\equiv \textrm{EW}_{\textrm{\FeII}}/\textrm{EW}_{\textrm{H}\beta}$, which is an Eigenvector 1 indicator.  Inclusion of \RFeII\ in predicting \lbol\ is significant at nearly the $3\sigma$ level and reduces the scatter and systematic offset of the luminosity residuals.  Typically, \oiii\ bolometric corrections are adopted for Type 2 sources where the quasar continuum is not observed and in these cases, \RFeII\ cannot be measured.  We searched for an alternative measure of Eigenvector 1 that could be measured in the optical spectra of Type 2 sources but were unable to identify one.  Thus, the main contribution of this work is to present an improved \oiii\ bolometric correction based on measured bolometric luminosities and highlight the Eigenvector 1 dependence of the correction in Type 1 sources.
\end{abstract}


\section{INTRODUCTION}
Active galactic nuclei (AGN) are among the most energetic phenomena in the universe, powered by accretion of gas onto supermassive black holes in the centers of distant galaxies.  The total energy output of an AGN is a fundamental quantity of significant interest.  However, because they emit their substantial energy over the entire electromagnetic spectrum, quantifying the bolometric luminosity is a particularly expensive task in terms of telescope time and is fraught with many pitfalls \citep[for a discussion, see][]{Runnoe12a}.  Fortunately, the spectral energy distributions \citep[SEDs, e.g.,][]{wilkes04} of AGN are roughly similar in shape and a bolometric correction, which sets the normalization of an average SED, will yield a reasonably accurate measure of the bolometric luminosity based on observations in a limited wavelength regime.

Recently, we presented bolometric corrections based on monochromatic luminosities in the optical/ultraviolet (UV), X-ray \citep{Runnoe12a}, and infrared \citep[IR,][]{Runnoe12b} based on the SED sample of \citet{Shang11}.  These are appropriate for estimating bolometric luminosity in Type 1 objects, where the continuum and broad emission lines are evident in the object's spectrum.  However, for Type 2 objects observed in the optical another approach is needed.  In Type 2 objects, the optical/UV continuum from the AGN is hidden from view, which may be caused by an obscuring torus \citep[e.g.,][]{Antonucci93} or by material farther out in the host galaxy \citep[e.g.,][]{DiPompeo15a,Malkan98}. As a result, bolometric corrections which rely on the observation of an optical continuum luminosity are not indicative of the bolometric luminosity of the AGN.  Instead, bolometric corrections in Type 2 objects are typically based on observations of the luminosity of the narrow emissions lines, which are thought to be emitted from gas at larger size scales and appear in the spectra of both Type 1 and 2 AGN.  

The forbidden $[\textrm{O}\,\textsc{iii}]~\lambda5007$ line is the most common narrow-line bolometric luminosity indicator. The utility of this line lies in its strength and low contamination due to emission from photoionized gas in star forming regions (e.g., compared the $[\textrm{O}\,\textsc{ii}]~\lambda3727$ line).  \oiii-based bolometric corrections have been derived by many authors \citep[see][for a review]{Heckman14} and are widely used.  The most commonly adopted \oiii\ bolometric correction in the literature, derived by \citet{Heckman04}, is $\textrm{L}_\textrm{iso}/\textrm{L}_{[\textrm{O}\,\textsc{iii}]} = 3500$.  This was derived for a sample of Type 1 sources using the observed (i.e. not extinction corrected) \oiii\ luminosities.  Bolometric luminosities were not measured for the sample, so they determined the ratio between the \oiii\ luminosity and the continuum at 5100~\AA, and then adopted the \citet{Marconi04} bolometric correction.  \citet{Kauffmann09} suggest that the \citet{Heckman04} \oiii\ bolometric correction should be adjusted for extinction in the narrow-line region (NLR).  They argue that Seyfert galaxy NLRs suffer $1.5-2$~mag extinction based on the Balmer decrement \citep{Kewley06}, and thus the bolometric correction for extinction corrected \oiii\ luminosity should be in the range $600-800$.  \citet{LaMassa09} measure a bolometric correction of $\textrm{L}_\textrm{iso}/\textrm{L}_{[\textrm{O}\,\textsc{iii}]} = 700$ in a sample of Type 2 sources with \oiii\ luminosities corrected for extinction based on the Balmer decrement and a Milky Way extinction curve.  In this sample, bolometric luminosity was determined by comparing the \oiii\ to the mid-IR continuum luminosity and then adopting the bolometric correction determined from the clumpy torus models of \citet{Nenkova08}.  Additionally, \oiii\ bolometric corrections are luminosity dependent \citep{Netzer06}, and \citet{Netzer09} further demonstrates that they depend on the ionization state of the gas, which can be estimated by including a measure of the $[\textrm{O}\,\textsc{i}]~\lambda6363$ line.  This adjustment is physically motivated, but in practice requires a wider wavelength range and brighter sources, since $[\textrm{O}\,\textsc{i}]$ is a weak line.

Notably, the strength of the \oiii\ emission line is not uniquely determined by the bolometric luminosity of an AGN.  That is, at a given bolometric luminosity there is a spread in the strength of the \oiii\ line primarily due to what is known as ``Eigenvector 1" \citep[EV1,][]{Boroson92}.  EV1 is an empirical description of the spectral properties of AGN, and \oiii\ is a key player where the range of EV1 properties is often characterized by the anti-correlation between the equivalent widths (EW) of \oiii\ and optical \FeII. The physical driver behind EV1 is not completely known, although the Eddington fraction ($\textrm{L}/\textrm{L}_{\textrm{Edd}}$) correlates strongly with EV1 indicators \citep[][but see also \citealt{Runnoe13b}]{Boroson02,Marziani96,Shen14}.  
	
The goal of this work is to derive a new set of \oiii\ bolometric corrections as an addition to the IR, optical, UV, and X-ray suite already in the literature for the \citet{Shang11} SEDs.  This sample is unique because the optical-UV data were taken quasi-simultaneously, which permits the construction of some of the most reliable big blue bump shapes in the quasar literature and makes the sample ideal for this work.  The main contributions of this work are to present \oiii\ bolometric corrections based on measured bolometric luminosities (none of the works described above have measured \lbol) and to demonstrate their dependence on EV1.  We describe the sample and data used for this work, including the details of the relevant spectral properties, in Section~\ref{sec:sample}.  In Section~\ref{sec:analysis} we present the \oiii\ bolometric corrections and correct them for the effect of EV1.  We compare our results to previous work in Section~\ref{sec:discussion} and summarize them in Section~\ref{sec:summary}.

We adopt a cosmology with $\textrm{H}_\textrm{0} = 70$ km s$^{-1}$ Mpc$^{-1}$, $\Omega_{\Lambda} = 0.7$, and $\Omega_{m} = 0.3$.  Note that we also differentiate between the bolometric luminosity calculated under the assumption of isotropy $(L_{\textrm{iso}})$ and true bolometric luminosity ($L_{\textrm{bol}}$), which likely differ.
	
\section{THE SAMPLE}
\label{sec:sample}
For this work we used the SED atlas of optically bright Type~1 quasars from \citet{Shang11}.  Out of the 85 objects in the catalog, 63 have complete wavelength coverage suitable for calculating bolometric luminosity.  Of these, 53 (32 of which are radio loud and 21 of which are radio quiet) also have coverage of the \oiii\ doublet allowing the derivation of an \oiii\ bolometric correction.  The redshifts of this subset of objects span the range $0.0345<z<1.0002$.  

The optical/UV data that facilitate this analysis were taken quasi-simultaneously with the {\it Hubble Space Telescope} and McDonald or Kitt Peak National Observatories.  \citet{Shang11} applied two corrections to these data before incorporating them in the SEDs.  The first was an extinction correction to remove the effects of dust in the Milky Way using the empirical mean extinction law of \citet{Cardelli89} and the dust maps of \citet{Schlegel98}.  The optical/UV spectra can also be affected by contamination from the quasar host galaxy light.  Because these are luminous quasars the contamination is typically weak, but a correction is still applied when extracting the spectra.

\citet{Runnoe12a} calculated the bolometric luminosities for the 63 objects that have complete wavelength coverage.  The SEDs were integrated from \SI{1}{\micro\metre} to 8~keV to obtain $L_\textrm{{iso}}$, the bolometric luminosity derived under the assumption of isotropy.  The gaps in the near-IR and extreme UV portions of the SEDs were handled by interpolating a log-log power-law spectrum between the data points on either side.  The final bolometric luminosities for the 53 sources with \oiii\ coverage are in the range log$(\textrm{L}_{\textrm{iso}}/\textrm{erg s}^{-1}) = 45.1-47.3$, with full details of the calculation given in \citet{Runnoe12a}.  Uncertainties on the bolometric luminosities are potentially large, but difficult to quantify because they depend on the unobserved emission in the extreme UV.  \citet{Runnoe12a} determined that the choice of model to interpolate over this region can introduce an uncertainty of 30\%, or 0.1303 when propagated into log$(\textrm{L}_{\textrm{iso}}/\textrm{erg s}^{-1})$, on average and we adopt this value.  

The \oiii\ luminosities and \RFeII\ were calculated based on the spectral decompositions from \citet{Tang12}.  They used the IRAF package \citep{Kriss94} to perform a $\chi^2$
\include{table} 
\noindent minimization that simultaneously fits all components in a model to the spectrum in the \Hb\ region.  The model spectrum included a power law to characterize the quasar continuum emission, an optical \FeII\ template \citep{Boroson92}, and multiple Gaussians (to which no physical meaning is attributed) to characterize the emission lines.  The narrow \Hb\ and \oiii\ were represented by one Gaussian each, and an additional two were allotted to characterize the broad \Hb\ emission, which may be asymmetric.  	

The measured spectral properties were determined from the best-fitting model components of the spectral decomposition.  The \oiii\ luminosities were calculated from the \oiii\ fluxes listed in table~4 of \citet{Tang12} and are in the range log$(\textrm{L}_{[\textrm{O}\,\textsc{iii}]}/\textrm{erg s}^{-1}) = 41.1 - 43.8$.  These are not corrected for intrinsic reddening.  In many cases we do have coverage of the H$\alpha$ and \Hb\ narrow emission lines that should allow us to measure the Balmer decrement and estimate reddening in the NLR.  However, we find that the narrow Balmer lines are usually too blended or too weak to yield a confident estimate of the extinction.  These are bright, blue quasars with strong Big Blue Bumps, so this it is unlikely that a large reddening correction is needed, but we revisit this issue and its implications in Section~\ref{sec:discussion}.  The optical \FeII\ equivalent width, which goes into \RFeII$\equiv \textrm{EW}_{\textrm{\FeII}}/\textrm{EW}_{\textrm{H}\beta}$, was calculated from the best-fitting \FeII\ template between 4478 and 5640~\AA\ using the local continuum at 4861~\AA.  The formal uncertainties on the spectral decomposition were determined by \citet{Tang12} and we propagate these into uncertainties on log$(\textrm{L}_{[\textrm{O}\,\textsc{iii}]}/\textrm{erg s}^{-1})$ and \RFeII.  The final measured properties for the sample are presented in Table~\ref{tab:prop}.	

\section{DERIVING [O III] BOLOMETRIC CORRECTIONS}
\label{sec:analysis}
We determined several versions of the \oiii\ bolometric correction, which we describe in this section.  Traditionally, bolometric corrections are characterized as	$\textrm{\lbol}=A\,\textrm{\Loiii}$, so we determined this correction first for comparison with values in the literature.  This approach assumes that all SEDs are intrinsically the same shape and differ only by normalization.  We later relaxed this assumption and explored bolometric corrections of the form $\textrm{log(\lbol)}=B+C\,\textrm{log(\Loiii)}$, as a potential avenue to better describe the data.  Finally, we also derived a bolometric correction that predicts the bolometric luminosity based on \Loiii\ and \RFeII, thereby accounting for the fact that EV1 may introduce scatter into the relationship between \Loiii\ and \lbol.  We investigated the applicability of separate corrections for radio-loud and radio-quiet sources.  However, the luminosity overlap was insufficient to characterize any differences or similarities, so we treated them together.
	 			
All fits in this paper were determined by analytic chi-squared minimization.  An ordinary least squares Y on X fit was appropriate in this case because our goal was to predict bolometric luminosity rather than determine the best-fitting underlying relationship \citep{Isobe90}.  Effectively, this means that only the uncertainties in bolometric luminosity entered into the fits because it was always the predicted quantity.  As described in Section~\ref{sec:sample}, these are potentially substantial and not well characterized.  \citet{Runnoe12a} determined that the interpolation over the gap in the extreme UV can introduce 30\% uncertainty on average, which we adopted as the uncertainty in \lbol.  We note that in comparison, uncertainties in \Loiii\ and \RFeII\ are negligible because the monochromatic fluxes are well known and the spectra have high signal-to-noise.  Uncertainties on the fit parameters were taken to be the standard deviation of the distribution for each coefficient determined by bootstrap sampling the bolometric luminosities, and re-calculating the fit $10^5$ times.  In all cases, we display equations for our best-fitting relationships with enough significant figures in order to reproduce the figures.	

\subsection{Traditional bolometric corrections}
\begin{figure}
	\centering
	\includegraphics[scale = .45]{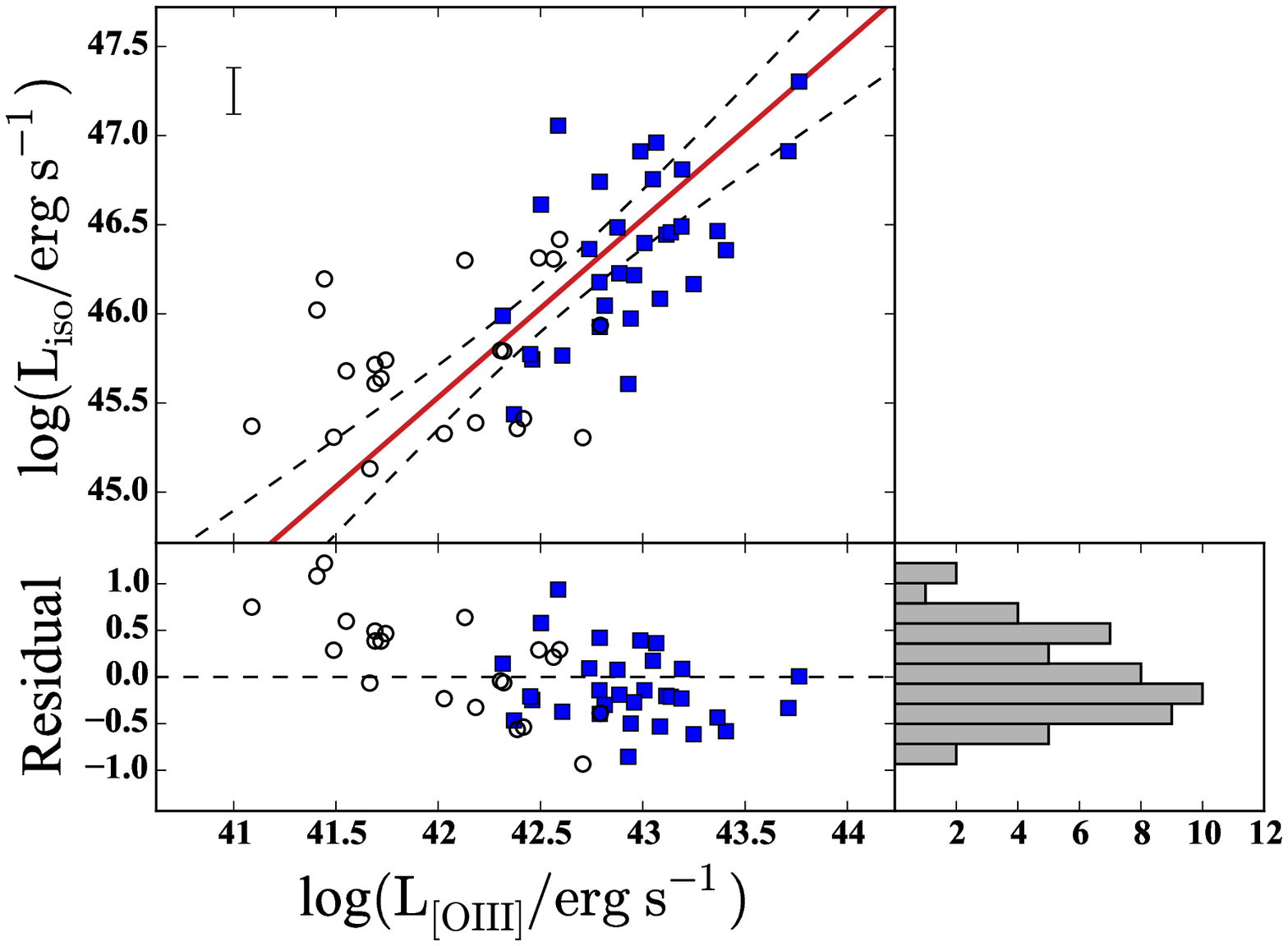}
	\caption {{\it Top}: log(\lbol) versus log(\Loiii). The red line represents a bolometric correction of the form $\textrm{\lbol} = A\,\textrm{\Loiii}$ and dashed lines indicate the 95\% confidence intervals.  Blue squares indicate radio-loud objects and open circles indicate radio-quiet objects.  A characteristic error bar for the data points is shown in the top left.  The uncertainty in \Loiii\ is small compared to \lbol\ and is not used in the fit, so we do not show it.  {\it Bottom left}: Residuals from the fit with a dashed line indicating where the predicted and observed bolometric luminosities are equal.  There is a clear trend in the residuals as a function of luminosity.  {\it Bottom right}: The distribution of the residuals derived from the scatter plot has a median of $-0.07$~dex and standard deviation $0.47$~dex. \label{fig:zerobc}}
\end{figure}

In keeping with convention, we found the ratio between \lbol\ and \Loiii\ by fitting a line with slope unity through log(\lbol) versus log(\Loiii).  This fit yielded the bolometric correction
\begin{equation}
\textrm{log}\left(\frac{\textrm{L}_{\textrm{iso}}}{\textrm{erg~s}^{-1}}\right) = \textrm{log}\left(\frac{\textrm{L}_{[\textrm{O}\,\textsc{iii}]}}{\textrm{erg~s}^{-1}}\right)+(3.532\pm 0.059),
\end{equation}
\noindent which is shown in Figure~\ref{fig:zerobc}.  The histogram of residuals, shown in the bottom right panel of Figure~\ref{fig:zerobc}, has a median of $-0.07$~dex and standard deviation of 0.47~dex.  We noted that the lower-luminosity objects tend to be scattered above the best-fitting line, whereas most of the higher-luminosity objects lie below the line.  Visually, it is evident that this fit does not completely capture these data.  Nevertheless, in linear space this can easily be compared to values in the literature:
\begin{equation}
\label{eqn:zerobc}
{L_{\textrm{iso}}}/L_{[\textrm{O}\,\textsc{iii}]} = 3400.
\end{equation}

\subsection{An improved [O\,{\sc iii}] bolometric correction}	
\begin{figure}
	\centering
	\includegraphics[scale = .45]{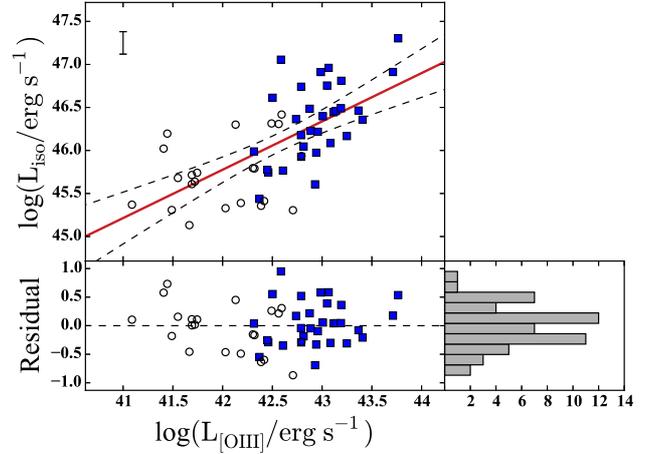}
	\caption {{\it Top}: log(\lbol) versus log(\Loiii).  The red line represents a bolometric correction of the form \lbol = B + C \Loiii and dashed lines indicate the 95\% confidence intervals.  Blue squares indicate radio-loud objects and open circles indicate radio-quiet objects.  A characteristic error bar for the data points is shown in the top left.  The uncertainty in \Loiii\ is small compared to \lbol\ and is not used in the fit, so we do not show it. {\it Bottom left}: Residuals from the fit with a dashed line indicating where the predicted and observed bolometric luminosities are equal. There is a slight trend in the residuals as a function of luminosity. {\it Bottom right}: The distribution of the residuals derived from the scatter plot has a median of $0.02$~dex and standard deviation $0.39$~dex. \label{fig:nzerobc}}		
\end{figure}

\citet{Nemmen10} showed, and \citet{Runnoe12a} confirmed, that bolometric corrections to the optical and UV continuum luminosities are better characterized by an expression with non-unity slope in logarithmic space.  Motivated by the trend in the residuals seen in the bottom panel of Figure~\ref{fig:zerobc} and the knowledge that there is significant diversity in the shape of quasar SEDs \citep{Shang05}, we explored this possibility for \oiii.  We derived the bolometric correction:
\begin{eqnarray}
\label{eqn:nzerobc}
\nonumber \textrm{log}\left(\frac{\textrm{L}_{\textrm{iso}}}{\textrm{erg~s}^{-1}}\right) &=& (0.5617\pm 0.0978)\,\textrm{log}\left(\frac{\textrm{L}_{[\textrm{O}\,\textsc{iii}]}}{\textrm{erg~s}^{-1}}\right) \\
&+&(22.186\pm 4.164),
\end{eqnarray}
\noindent shown in Figure~\ref{fig:nzerobc}.  The measurements now appear evenly distributed around the best-fitting line and the residuals are more well-behaved.  The distribution of residuals has median $0.02$~dex and standard deviation $0.39$~dex, a moderate improvement.
 
Visually, it appears that the \oiii\ bolometric correction may depend on radio loudness or luminosity.  Specifically, in Figures~\ref{fig:zerobc} and \ref{fig:nzerobc}, the high-luminosity radio-loud and low-luminosity radio-quiet sources appear to follow trends with slightly different slopes.  Because there are so few radio-loud and radio-quiet sources with similar luminosities, we were unable to determine whether there are differences due to radio loudness at a given luminosity.  That is, we could not separate the effects of radio loudness from potential luminosity effects.  Because there is no evidence that directly indicates radio-loud and radio-quiet sources should be treated differently and a luminosity dependence is expected from the literature \citep[e.g.,][]{Netzer06}, we chose to address the latter.  We therefore explored the possibility that higher orders in log(\Loiii) might better predict log(\lbol) by fitting the data with a function of the form $\textrm{log(\lbol)}=B+C\,\textrm{log(\Loiii)}+D\,\textrm{log(\Loiii)}^2$.  To balance the fact that a function with more free parameters will always perform better in a chi-squared test, we used the Bayesian Information Criterion \citep{Schwartz78} to compare the performance of this fit compared to the one in Equation~\ref{eqn:nzerobc}.  We found that the improvement in the fit was not sufficient to warrant the addition of more free parameters, indicating that the data do not support the derivation of a luminosity-dependent bolometric correction.  Taken together, this indicates that the bolometric correction in Equation~\ref{eqn:nzerobc} is the best narrow-line predictor of \lbol\ that we can derive for this sample.

\subsection{Accounting for the diversity of quasar spectra}
\begin{figure*}
\begin{subfigure}[!b]{8cm}
  \centering
  \includegraphics[scale=0.4]{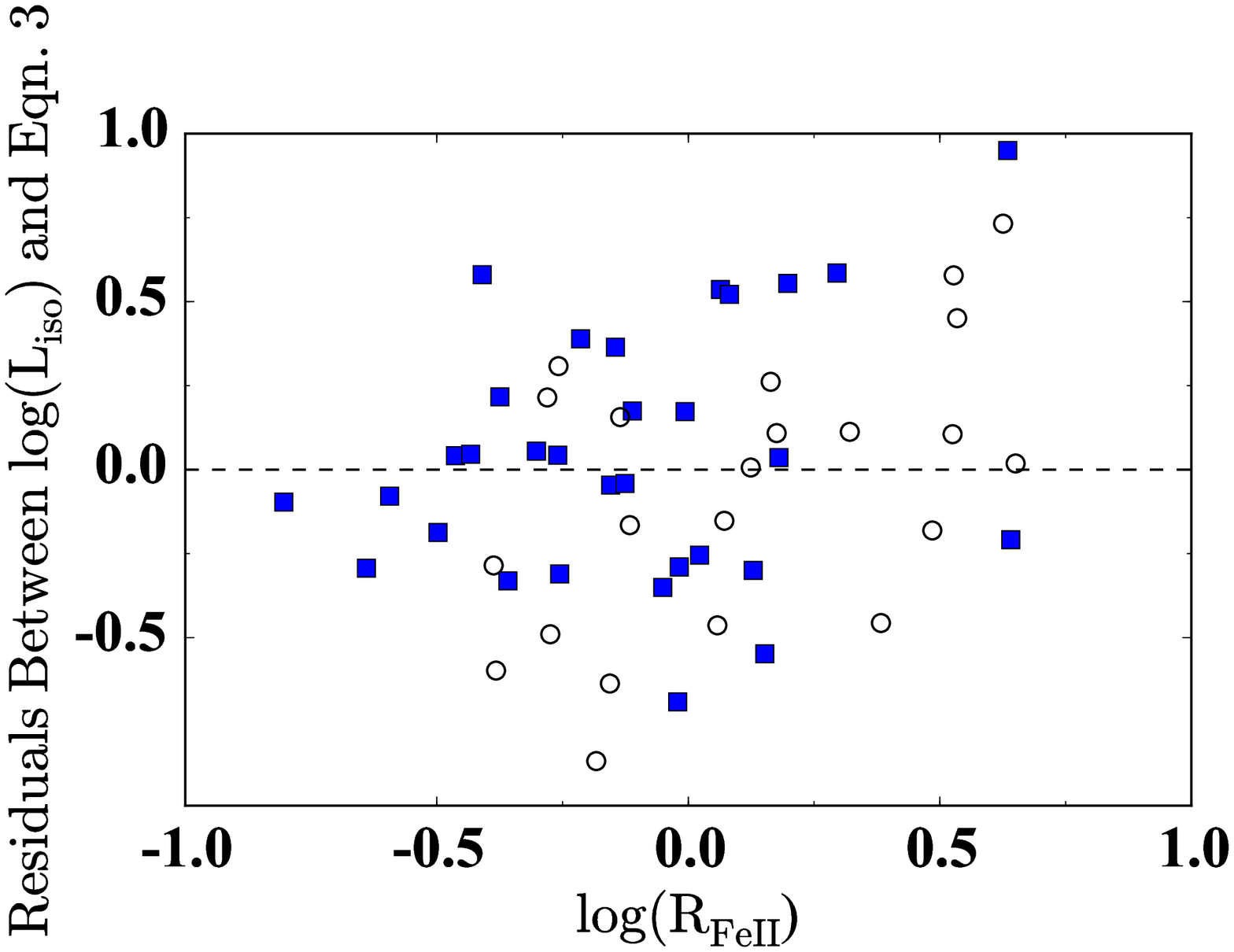}
\end{subfigure}
\hspace{0.1cm}
\begin{subfigure}[!b]{8cm}
	\centering
	\includegraphics[scale=0.4]{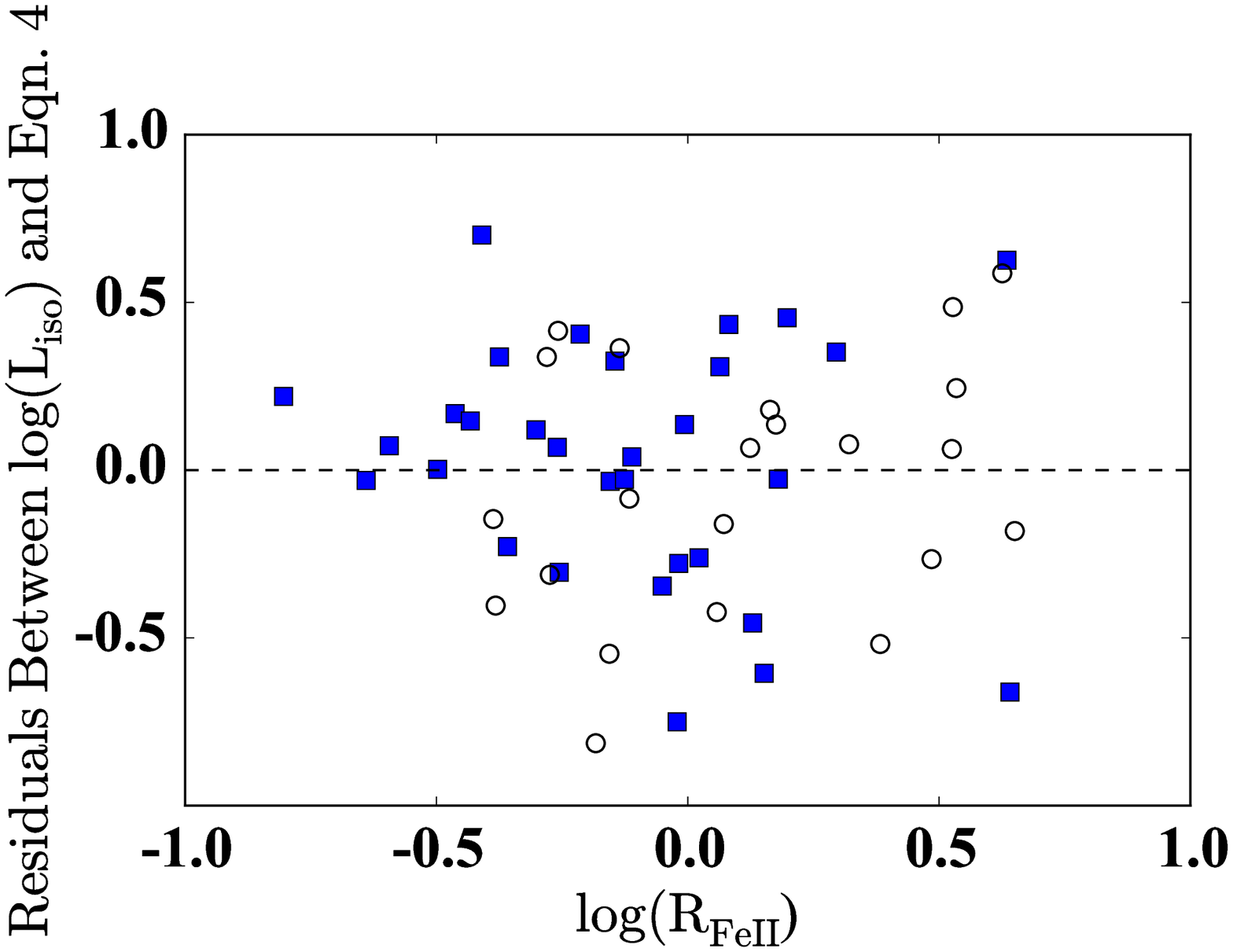}
\end{subfigure}
\hspace{0.1cm}
\begin{subfigure}[!b]{8cm}
  \centering
  \includegraphics[scale=0.4]{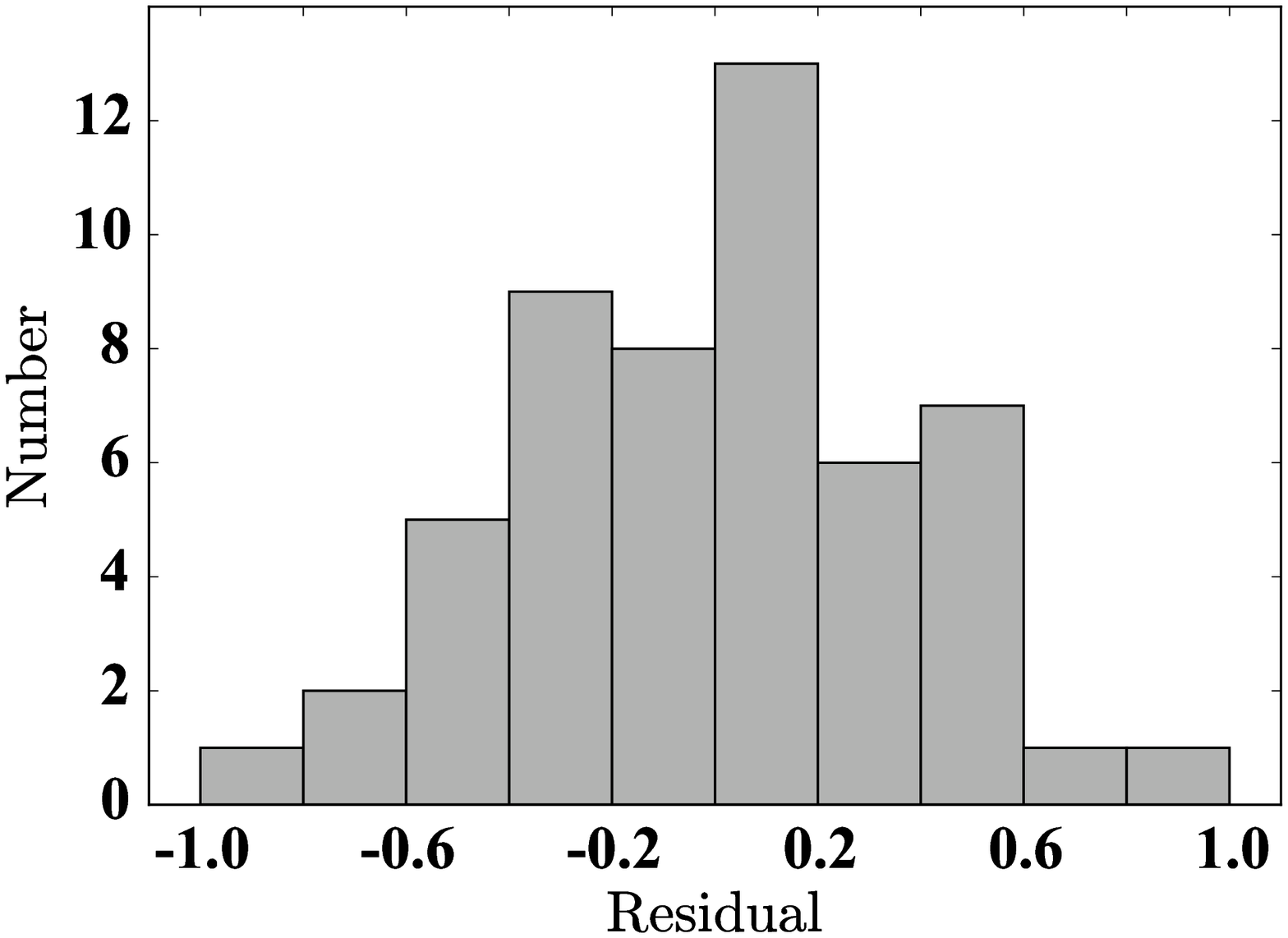}
\end{subfigure}
\hspace{0.1cm}
\begin{subfigure}[!b]{8cm}
  \centering
  \includegraphics[scale=0.4]{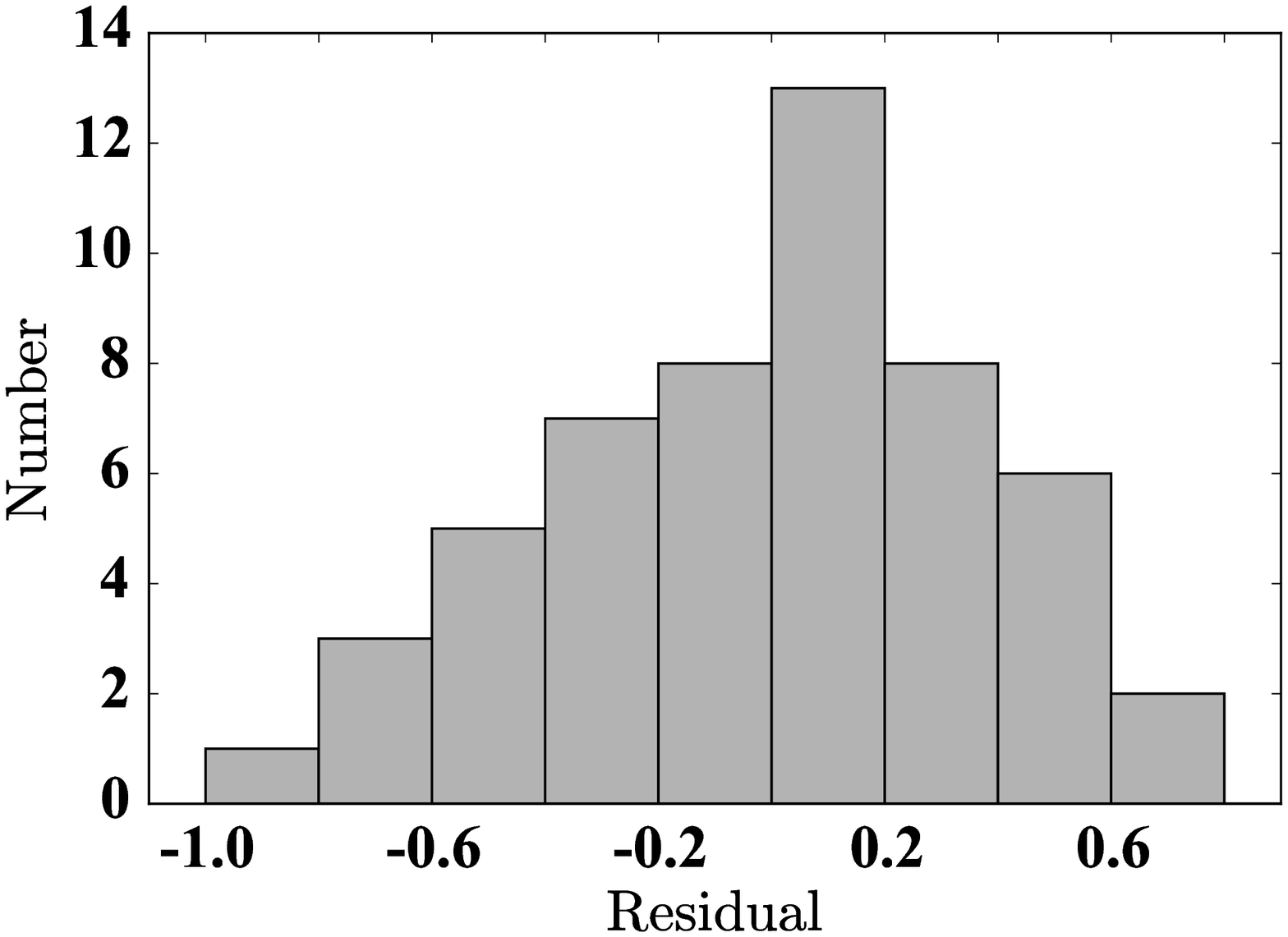}
\end{subfigure}
\caption{{\it Left}: The residuals between the measured bolometric luminosity and the calculated values from Equation~\ref{eqn:nzerobc}.  The top panel shows the residuals versus log(\RFeII) and the bottom panel shows their distribution.  The trend in the bolometric luminosity residuals versus \RFeII\ has a Spearman Rank correlation coefficient of 0.259 and associated probability of 0.061.  The distribution of luminosity residuals has median 0.02~dex and standard deviation 0.39~dex.
{\it Right}:  Same as the left panels, but the bolometric luminosity residuals are calculated using Equation~\ref{eqn:ev1bc} and accounting for EV1.  There is no longer a significant correlation with \RFeII\ and the distribution of residuals is centered near zero with reduced width.  The distribution of residuals has median 0.047~dex and standard deviation 0.36~dex.  Blue squares indicate radio-loud objects and open circles indicate radio-quiet objects.  \label{fig:ev1resid}}
\end{figure*}

Because the \oiii\ line in quasar spectra is not solely determined by the source luminosity, we expected that we could improve our ability to predict \lbol\ by including some indication of an object's position along EV1.  To explore this possibility, we looked at the residuals between the observed \lbol\ and those calculated from Equation~\ref{eqn:nzerobc} versus \RFeII, as shown in the left panels of Figure~\ref{fig:ev1resid}.  The Spearman Rank correlation coefficient and associated probability of finding the observed distribution of points by chance are 0.259 and 0.061, respectively.  This confirms that there is a weak but marginally significant correlation between the bolometric luminosity residuals and our EV1 indicator of choice, \RFeII.  We calculated a new bolometric correction with \oiii\ and \RFeII\ by applying a multiple linear regression model to both proxies. This gave:	
\begin{eqnarray}
\label{eqn:ev1bc}
\nonumber \textrm{log}\left(\frac{L_{\textrm{iso}}}{\textrm{erg~s}^{-1}}\right) &=& (0.7144\pm 0.1170)\,\textrm{log}\left(\frac{L_{[\textrm{O}\,\textsc{iii}]}}{\textrm{erg~s}^{-1}}\right) \\
\nonumber &+& (0.4838 \pm 0.2007)\,\textrm{log}(R_{\textrm{Fe}\,\textsc{ii}}) \\
&+& (15.702\pm 4.975).
\end{eqnarray}
We found that the inclusion of the \RFeII\ term in predicting \lbol\ is significant at nearly the 3$\sigma$ level.  In order to determine the effects of including \RFeII, we first revisited the residuals between log(\lbol) and the luminosities predicted by Equation~\ref{eqn:ev1bc} versus \RFeII\ (right panels of Figure~\ref{fig:ev1resid}).  The Spearman Rank correlation coefficient drops down to 0.005, and the probability of finding this distribution of points by chance is high at 0.970.  The distribution of residuals has median 0.047~dex and standard deviation 0.36~dex.

\begin{figure*}
\begin{subfigure}[!b]{8cm}
  \centering
  \includegraphics[scale=0.4]{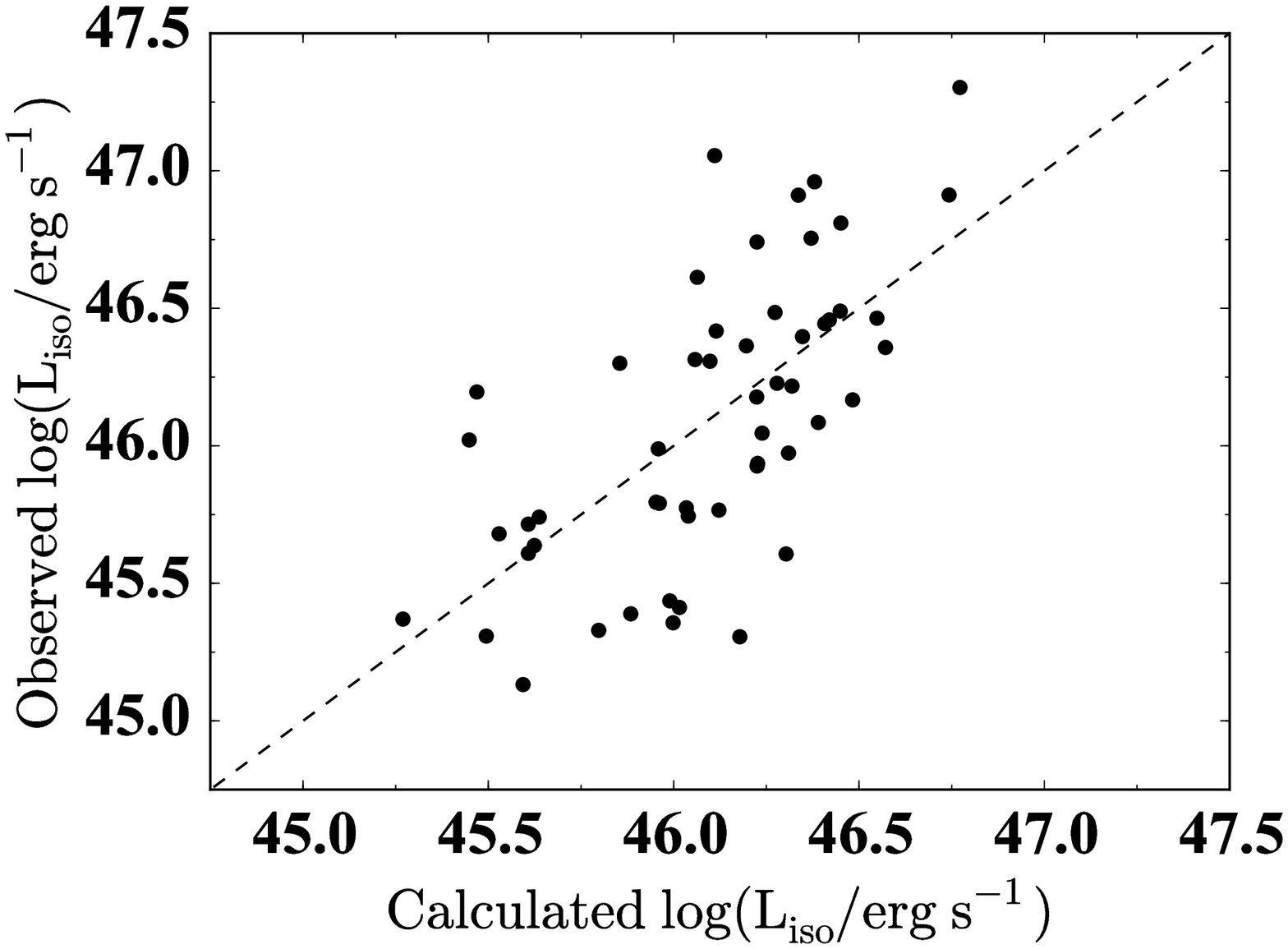}
\end{subfigure}
\hspace{0.1cm}
\begin{subfigure}[!b]{8cm}
	\centering
	\includegraphics[scale=0.4]{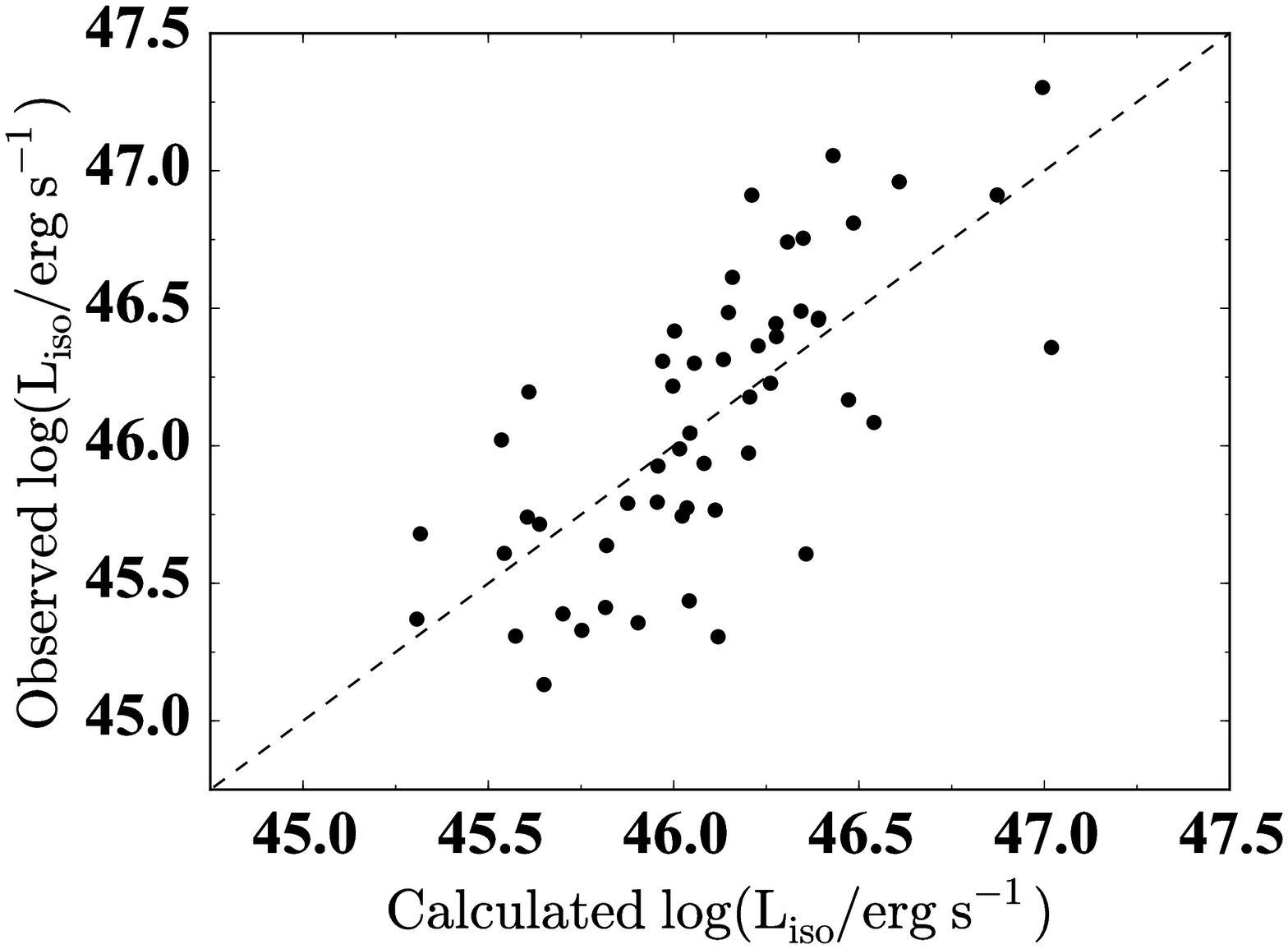}
\end{subfigure}
\caption{{\it Left}: The observed luminosities compared to those predicted based on \oiii\ alone by Equation~\ref{eqn:nzerobc}.  {\it Right}: The observed luminosities now compared to those predicted by Equation~\ref{eqn:ev1bc}, which accounts for EV1.  The dashed lines in each panel show where the luminosities are equal.  Including a term for \RFeII\ in order to account for EV1 improves the agreement with the measured luminosities.  \label{fig:ev1bc}}
\end{figure*}

To further visualize the improvement associated with adopting the \RFeII\ correction, in Figure~\ref{fig:ev1bc} we show the measured \lbol\ values against the predicted luminosities from Equation~\ref{eqn:ev1bc}.  The modest reduction in scatter is visually evident, as the data appear more centered around the one-to-one line.

\section{DISCUSSION}
\label{sec:discussion}
\begin{figure}
	\centering
	\includegraphics[scale = .4]{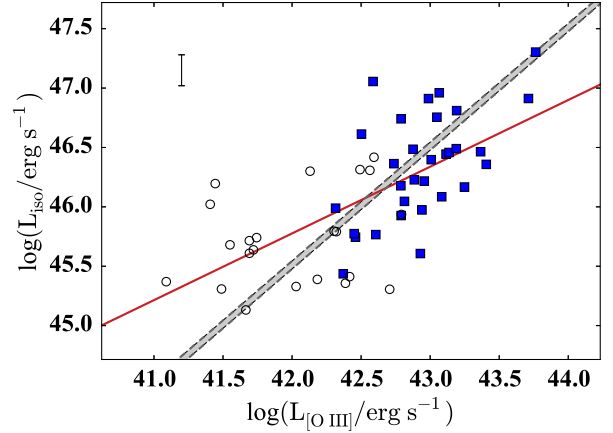}
	\caption {log(\lbol) versus log(\Loiii).  The red line represents our best \oiii-only correction from Equation~\ref{eqn:nzerobc}.  The light gray shaded area represents the range of other bolometric corrections found in the literature of the form $\textrm{\lbol} = A \textrm{\Loiii}$ for observed \Loiii.  With a value of 3400, our correction falls near the top of this range.  Radio-loud points are displayed as blue squares and radio-quiet points are open circles. A characteristic error bar for the data points is shown in the top left.  The uncertainty in \Loiii\ is small compared to \lbol\ and is not used in the fit, so we do not show it. \label{fig:bccomp}}		
\end{figure}

We find good agreement between our bolometric correction described in Equation~\ref{eqn:zerobc} and those in the literature, which we show in Figure~\ref{fig:bccomp}.  The agreement with the commonly used correction from \citet{Heckman04} is particularly good, but perhaps not surprising, as both corrections were measured for Type 1 sources with \oiii\ luminosities uncorrected for extinction.  Compared to the sample of \citet{Heckman04}, our SED sample has higher luminosity with substantial overlap ($41.1<\textrm{log(\Loiii/erg s}^{-1})< 43.8$ versus $40.1 <\textrm{log(\Loiii/erg s}^{-1})< 42.6$).  The \citet{Heckman04} value is anchored to the \citet{Marconi04} bolometric correction to the optical continuum luminosity, rather than based on measured bolometric luminosities in their sample.  The \citet{Marconi04} corrections are determined from a model SED that matches the \citet{elvis94} observed SEDs well in the optical/UV but is moderately weaker in X-rays.  As we do, they exclude the reprocessed IR emission when integrating the SED so as not to double count photons, although they have a slightly higher upper limit of integration.  

If the \oiii\ luminosity is attenuated due to dust, as some authors have suggested, these \oiii\ bolometric corrections will need to be adjusted.  There is evidence to suggest that Type 2 sources may suffer more dust extinction (by $1-2$~mag) than Type 1s \citep{Diamond-Stanic09}, but that the \oiii\ line can be corrected to within a factor of 3 for this effect \citep{Wild11}.  Physically, the NLR is likely stratified and, depending on the density profile, there can be compact emitting regions that may be preferentially obscured by a dusty torus in Type 2 sources.  An extinction correction appears to be necessary in Type 2 sources, and may be required independent of optical spectral classification (i.e. Type 1 versus Type 2) because dust extinction in the NLR would occur on large scales exterior to a dusty obscuring structure.  If we were to assume, as \citet{Kauffmann09} do, that the Balmer decrements measured by \citet{Kewley06} for emission-line galaxies with AGN-like line ratios imply $1.5-2$~mag of extinction in the NLR, we would obtain smaller bolometric corrections consistent with \citet{LaMassa09}.  While we do have H$\alpha$ coverage in many of our sources, the narrow lines are often not visually distinguishable and the spectral decomposition is not sufficiently detailed to characterize them accurately.  Furthermore, since measuring the amount of extinction and recovering the intrinsic \oiii\ luminosity includes many uncertainties in practice and so we do not apply any extinction corrections to the \oiii\ luminosity here.
	
\begin{figure*}
\begin{subfigure}[!b]{8cm}
  \centering
    \includegraphics[scale=0.4]{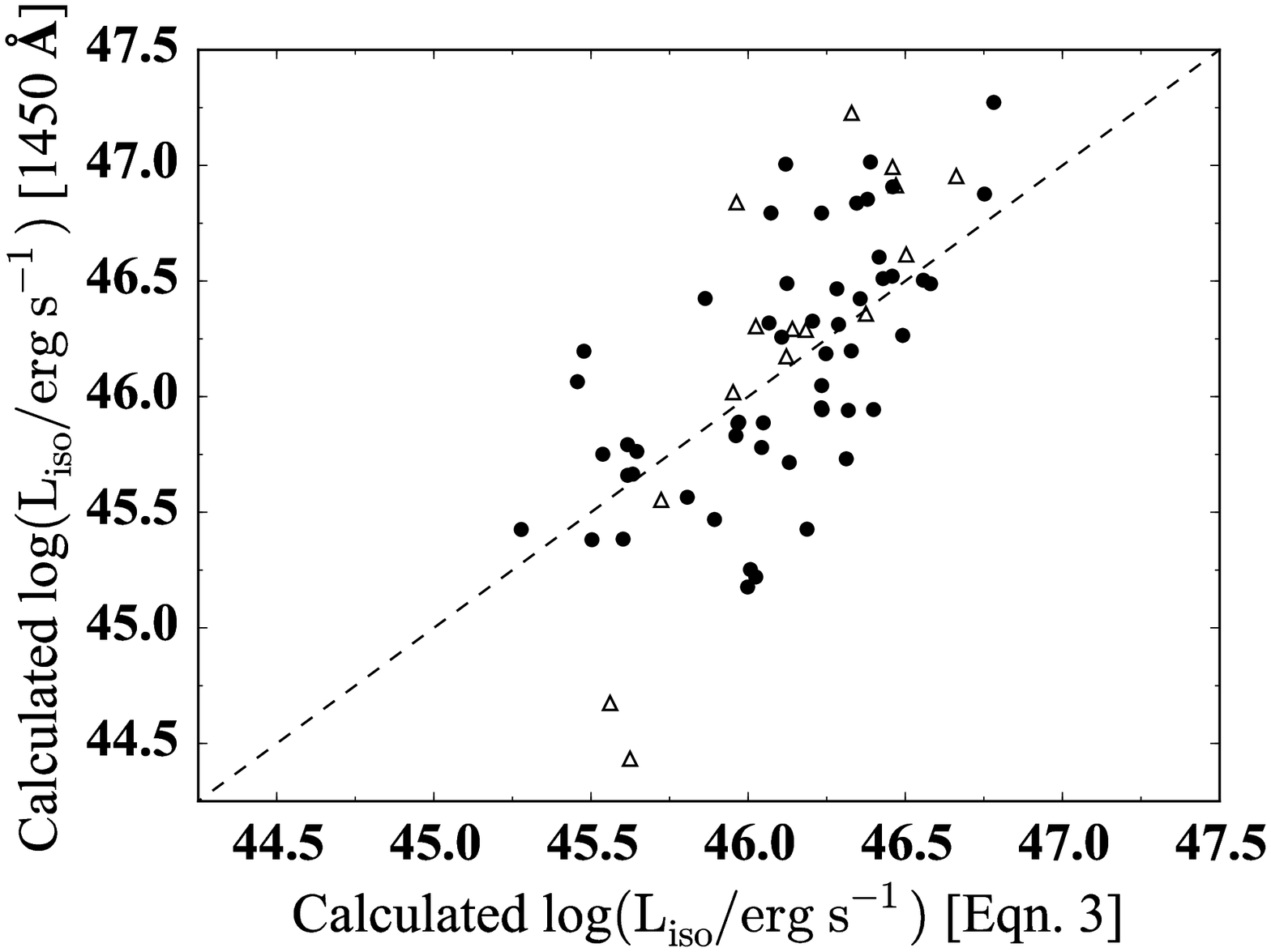}
\end{subfigure}
\hspace{0.1cm}
\begin{subfigure}[!b]{8cm}
	\centering
	  \includegraphics[scale=0.4]{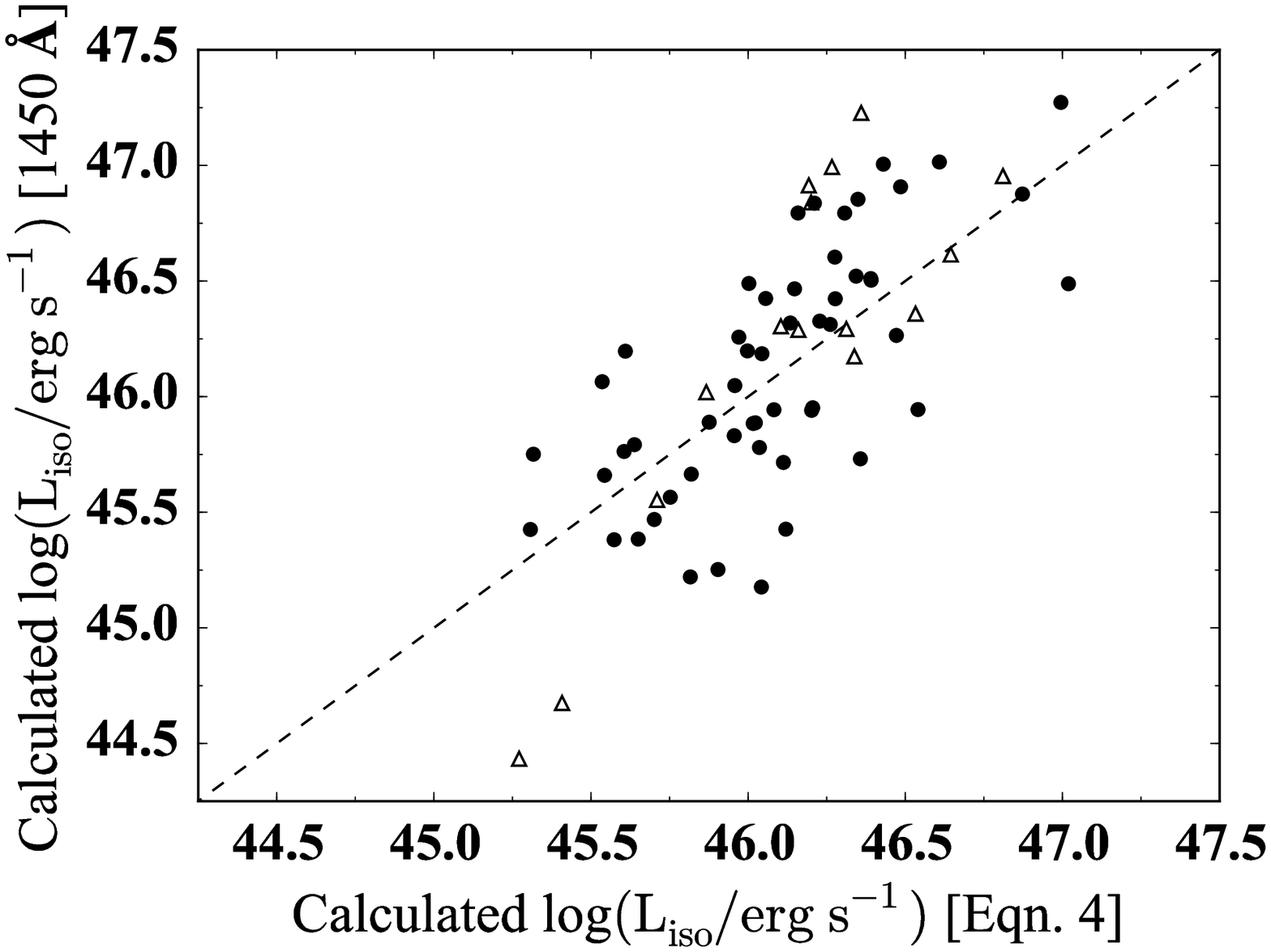}
\end{subfigure}
\hspace{0.1cm}
\begin{subfigure}[!b]{8cm}
  \centering
  \includegraphics[scale=0.4]{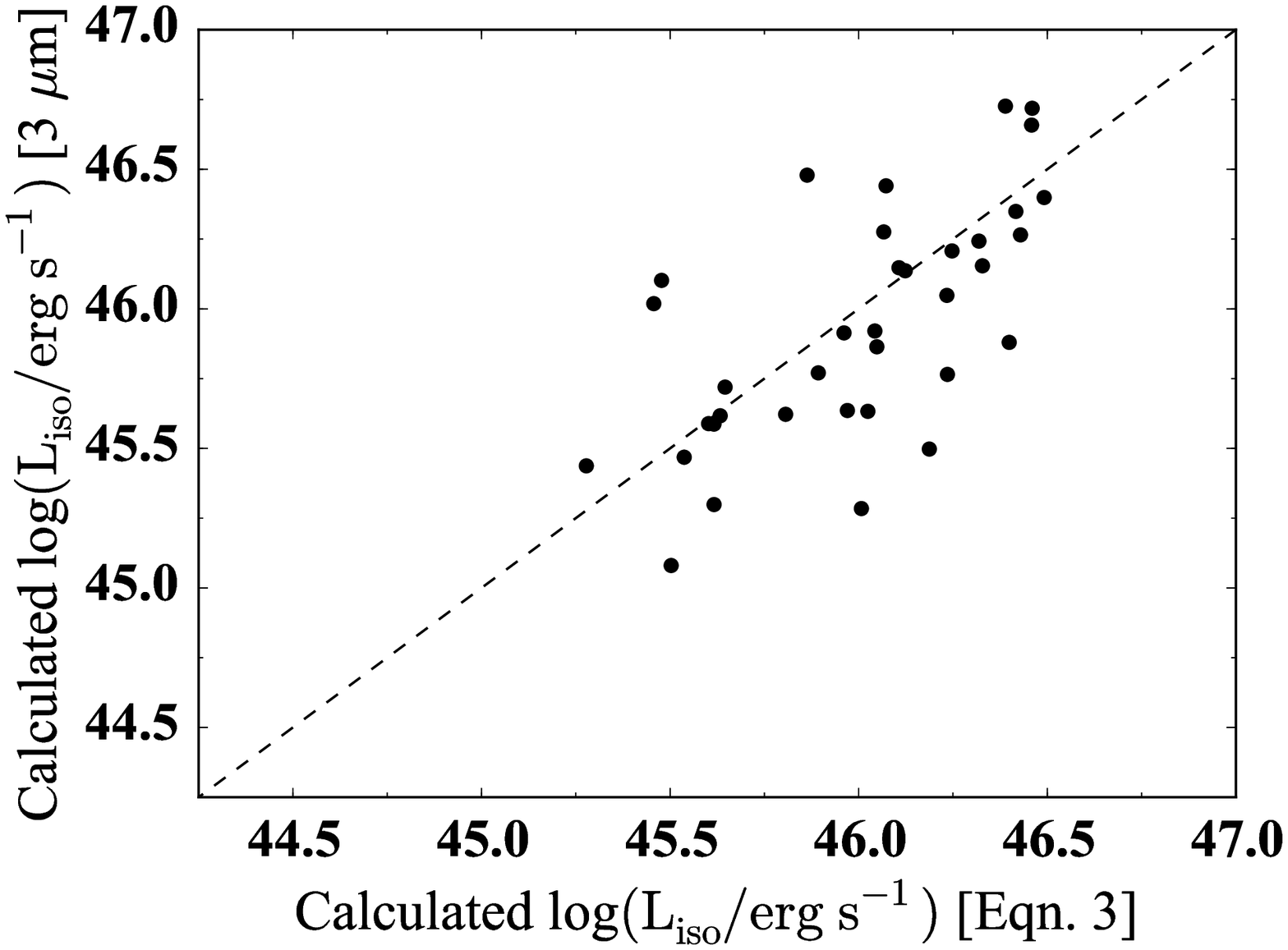}
\end{subfigure}
\hspace{0.1cm}
\begin{subfigure}[!b]{8cm}
  \centering
	\includegraphics[scale=0.4]{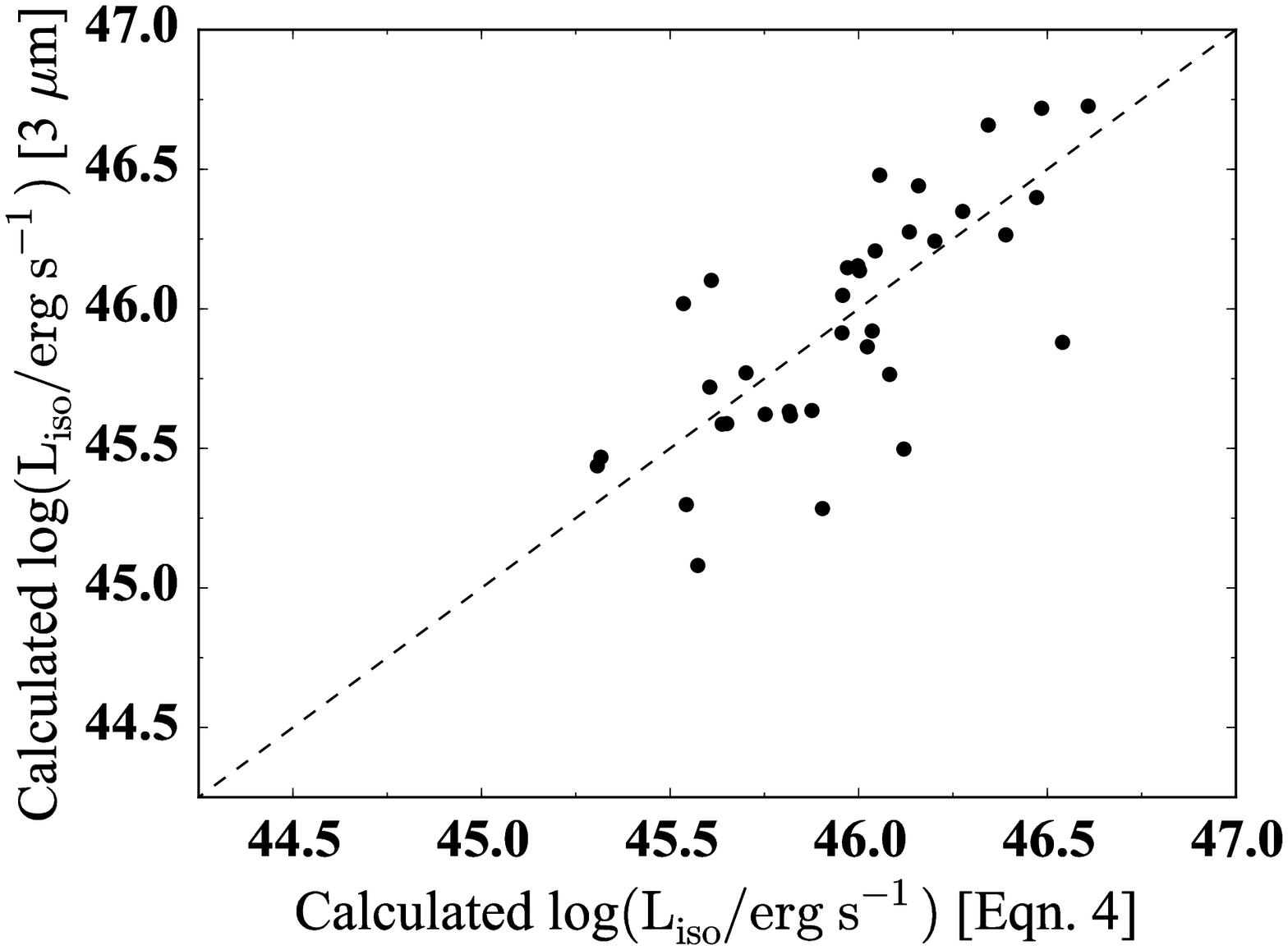}
\end{subfigure}
\caption{{\it Left}: Comparison between the bolometric luminosities calculated from the quasar continuum emission at 1450~\AA\ \citep[top panel,][]{Runnoe12a} and \SI{3}{\micro\metre} \citep[bottom panel,][]{Runnoe12b} to those calculated from Equation~\ref{eqn:nzerobc}.  The dashed line shows where the luminosities are equal and the open triangles indicate objects that were not used to derive the bolometric corrections.  The scatter is 0.44 and 0.32~dex for the 1450~\AA\ and \SI{3}{\micro\metre} corrections, respectively.  {\it Right}:  Same as the left column, but compared to luminosities calculated from our Equation~\ref{eqn:ev1bc} and thus accounting for the EV1 contribution to \oiii.  The scatter is reduced compared to the left panel at 0.40 and 0.29~dex for the 1450~\AA\ and \SI{3}{\micro\metre} corrections, respectively.  Additionally, the systematic trend that was apparent in the comparison with UV-based luminosities disappears. \label{fig:contcomp}}
\end{figure*}

We can also compare our \oiii\ bolometric corrections to the 1450~\AA\ and \SI{3}{\micro\metre} continuum corrections derived from this SED sample by \citet{Runnoe12a} and \citet{Runnoe12b}, respectively.  The left column of Figure~\ref{fig:contcomp} shows this comparison for luminosities derived from our Equation~\ref{eqn:nzerobc}.  For the UV, we include in the comparison 14 objects from the \citet{Shang11} atlas that were not used to determine the bolometric corrections because they did not have appropriate data coverage.  The agreement is reasonably good, with scatter of 0.44 and 0.32 dex for the UV and IR, respectively.  However, visual inspection shows a systematic trend between luminosities calculated from the \oiii\ and UV bolometric corrections.  This is the effect of ignoring EV1.  The right column of Figure~\ref{fig:contcomp} shows the comparison with luminosities derived from our Equation~\ref{eqn:ev1bc}; the scatter is reduced to 0.40 and 0.29~dex for the UV and IR, respectively, and the systematic trend is resolved.

The morphology of the \oiii\ emitting region is known to vary among AGN and will play a role in the applicability of \oiii\ bolometric corrections.  Because the size of the NLR as measured by \oiii\ emission scales with the source luminosity \citep{Bennert02,Hainline13}, this suggests that the corrections derived here should be applied only to sources of similar luminosity.  Indeed, \oiii\ bolometric corrections are known to be luminosity dependent \citep{Netzer06,LaMassa10,Hainline13,Heckman14}, although the data in this work were insufficient to demonstrate this behavior.  At low luminosity the NLR is smaller and a bigger fraction of the \oiii\ luminosity may be blocked by a dusty torus.  Consistent with this idea, \citet{Kauffmann09} consider high and low-luminosity sources separately and identify a smaller bolometric correction ($300-600$ for extinction corrected \oiii) for low-luminosity AGN.  At very high luminosity the NLR becomes very large \citep[e.g.,][]{Hainline13,Zakamska16}.  In these conditions, the NLR size-luminosity relationship flattens out \citep{Hainline13,Hainline14}, and the \oiii\ luminosities may become saturated as the AGN photoionizes the entire interstellar medium of the host galaxy \citep[e.g.,][]{Hainline13}.

The primary utility of \oiii\ bolometric corrections is to determine the bolometric luminosity of Type 2 objects, so we must examine the assumption that Type 1 and Type 2 sources require the same bolometric correction.  Bolometric luminosity is obscured in Type 2 sources and therefore cannot be measured, so the only possible approach here is to compare the SEDs of Type 1 and Type 2 sources in wavelength regions that are not prone to extinction (i.e. the radio, IR, and hard X-rays).  As \citet{Heckman04} review in detail, no differences are found in these regimes based on optical spectral classification, suggesting that it is fair to apply bolometric corrections derived from Type 1 samples to Type 2 sources.

A further consideration relevant for the inclusion of an \RFeII\ term in an \oiii\ bolometric correction is that the spectra of Type 2 sources lack the characteristic broad emission lines and non-stellar continuum of their Type 1 counterparts.  This means that the \RFeII\ quantity cannot be measured in such sources because it is an equivalent width ratio between the broad \Hb\ and broad optical \FeII\ emission.  An EV1 indicator that relies only on spectral features observed in Type 2 sources would certainly be very useful for this and similar applications and we sought to identify one.  Our approach was to generate composite spectra binned by \RFeII\ and visually inspect them to look for optical narrow line ratios that tracked the broad-line EV1 proxy.  We specifically targeted emission lines in the $3700-4700$~\AA\ range (e.g., \OIIdblt, \NeIIIdblt, \SIIw) that were likely to be covered along with the \oiii\ emission line in an optical spectrum and therefore useful for our bolometric corrections.  Unfortunately, the composites did not reveal any strong trends with a dynamic range suitable for our purposes.  The main issues are that, with the exception of \oiii, the optical narrow lines in quasars are typically weak and difficult to measure reliably.  In Type 1 sources, the very nature of EV1 works against us in identifying a narrow-line EV1 indicator because on one end of EV1 the broad \FeII\ emission is strong and the narrow lines are weak, making the measurements even more difficult.  While there may be narrow-line ratios that scale with EV1, careful sample selection will be required to tease them out and it is unlikely that they will have the dynamic range needed to adjust \oiii\ bolometric corrections.  As a result of all this, while we have shown that EV1 introduces scatter into the \lbol\ versus \Loiii\ relationship, the correction for this effect cannot be calculated in Type 2 objects and a systematic offset and increased scatter will be unavoidable.  The best that can be done in these cases is to be aware of the effect and, luckily, the size of this effect is relatively modest.
	
\section{SUMMARY}
\label{sec:summary}
In this work we presented \oiii\ bolometric corrections based on 53 optical bright, Type 1 quasars with $0.0345 < z < 1.0002$ and log$(\textrm{L}_{\textrm{iso}}/\textrm{erg s}^{-1}) = 45.1-47.3$, of which 32 are radio loud and 21 are radio quiet.  These constitute an improvement to the existing suite of IR, optical, UV, and X-ray bolometric corrections derived from the \citet{Shang11} SEDs.  We derived the following three corrections:
\begin{itemize}
\item A correction for comparison with the literature, equivalent to ${\textrm{L}_{\textrm{iso}}}/\textrm{L}_{[\textrm{O}\,\textsc{iii}]} = 3400$:
\begin{eqnarray}
\textrm{log}(L_{\textrm{iso}}) &=& \textrm{log}(L_{[\textrm{O}\,\textsc{iii}]})+(3.532\pm 0.059).
\end{eqnarray}
\item An improved bolometric correction that reduces the offset in comparison to measured bolometric luminosities:
\begin{eqnarray}
\nonumber\textrm{log}(L_{\textrm{iso}})  &=& (0.5617\pm 0.0978)\,\textrm{log}(L_{[\textrm{O}\,\textsc{iii}]}) \\
&+&(22.186\pm 4.164).
\end{eqnarray}
\item A bolometric correction that accounts for the EV1 contribution to the \oiii\ line:
\begin{eqnarray}
\nonumber \textrm{log}(L_{\textrm{iso}})  &=& (0.7144\pm 0.1170)\,\textrm{log}(L_{[\textrm{O}\,\textsc{iii}]}) \\
\nonumber &+& (0.4838 \pm 0.2007)\,\textrm{log}(R_{\textrm{Fe}\,\textsc{ii}}) \\
&+& (15.702\pm 4.975).
\end{eqnarray}
\end{itemize}

The main contributions of this work, compared to existing \oiii\ bolometric corrections in the literature, are the use of measured bolometric luminosities and the derivation of the correction in Equation~\ref{eqn:nzerobc} (second in the list above), which we recommend for use on Type 2 objects.  We additionally demonstrated the EV1 bias in \oiii\ bolometric corrections, significant at nearly the $3\sigma$ level, which introduces scatter into the \Loiii--\lbol\ relationship.  The size of the effect is as much as a factor of 3 in bolometric luminosity for extreme values of \RFeII\ (our chosen EV1 indicator), and correcting for it improves agreement with corrections derived at other wavelengths.  We were unable to identify an optical EV1 indicator that can be measured in Type 2 sources, so this effect may only be acknowledged but not corrected at this time.

\section*{ACKNOWLEDGMENTS}
AP would like to thank Stephanie Brown for a careful reading of the manuscript and JCR acknowledges Mike Eracleous for comments on the manuscript and helpful discussions during the preparation of this work.  The authors thank the anonymous referee for their careful reading of the manuscript and constructive comments.

\bibliographystyle{mnras}
\bibliography{bibFile}

\begin{thebibliography}{}
\makeatletter
\relax
\def\mn@urlcharsother{\let\do\@makeother \do\$\do\&\do\#\do\^\do\_\do\%\do\~}
\def\mn@doi{\begingroup\mn@urlcharsother \@ifnextchar [ {\mn@doi@}
  {\mn@doi@[]}}
\def\mn@doi@[#1]#2{\def\@tempa{#1}\ifx\@tempa\@empty \href
  {http://dx.doi.org/#2} {doi:#2}\else \href {http://dx.doi.org/#2} {#1}\fi
  \endgroup}
\def\mn@eprint#1#2{\mn@eprint@#1:#2::\@nil}
\def\mn@eprint@arXiv#1{\href {http://arxiv.org/abs/#1} {{\tt arXiv:#1}}}
\def\mn@eprint@dblp#1{\href {http://dblp.uni-trier.de/rec/bibtex/#1.xml}
  {dblp:#1}}
\def\mn@eprint@#1:#2:#3:#4\@nil{\def\@tempa {#1}\def\@tempb {#2}\def\@tempc
  {#3}\ifx \@tempc \@empty \let \@tempc \@tempb \let \@tempb \@tempa \fi \ifx
  \@tempb \@empty \def\@tempb {arXiv}\fi \@ifundefined
  {mn@eprint@\@tempb}{\@tempb:\@tempc}{\expandafter \expandafter \csname
  mn@eprint@\@tempb\endcsname \expandafter{\@tempc}}}

\bibitem[\protect\citeauthoryear{{Antonucci}}{{Antonucci}}{1993}]{Antonucci93}
{Antonucci} R.,  1993, \mn@doi [\araa] {10.1146/annurev.aa.31.090193.002353},
  \href {http://adsabs.harvard.edu/abs/1993ARA%26A..31..473A} {31, 473}

\bibitem[\protect\citeauthoryear{{Bennert}, {Falcke}, {Schulz}, {Wilson}  \&
  {Wills}}{{Bennert} et~al.}{2002}]{Bennert02}
{Bennert} N.,  {Falcke} H.,  {Schulz} H.,  {Wilson} A.~S.,   {Wills} B.~J.,
  2002, \mn@doi [\apjl] {10.1086/342420}, \href
  {http://adsabs.harvard.edu/abs/2002ApJ...574L.105B} {574, L105}

\bibitem[\protect\citeauthoryear{{Boroson}}{{Boroson}}{2002}]{Boroson02}
{Boroson} T.~A.,  2002, \mn@doi [\apj] {10.1086/324486}, \href
  {http://adsabs.harvard.edu/abs/2002ApJ...565...78B} {565, 78}

\bibitem[\protect\citeauthoryear{{Boroson} \& {Green}}{{Boroson} \&
  {Green}}{1992}]{Boroson92}
{Boroson} T.~A.,  {Green} R.~F.,  1992, \mn@doi [\apjs] {10.1086/191661}, \href
  {http://adsabs.harvard.edu/abs/1992ApJS...80..109B} {80, 109}

\bibitem[\protect\citeauthoryear{{Cardelli}, {Clayton}  \& {Mathis}}{{Cardelli}
  et~al.}{1989}]{Cardelli89}
{Cardelli} J.~A.,  {Clayton} G.~C.,   {Mathis} J.~S.,  1989, \mn@doi [\apj]
  {10.1086/167900}, \href {http://adsabs.harvard.edu/abs/1989ApJ...345..245C}
  {345, 245}

\bibitem[\protect\citeauthoryear{{DiPompeo}, {Myers}, {Hickox}, {Geach},
  {Holder}, {Hainline}  \& {Hall}}{{DiPompeo} et~al.}{2015}]{DiPompeo15a}
{DiPompeo} M.~A.,  {Myers} A.~D.,  {Hickox} R.~C.,  {Geach} J.~E.,  {Holder}
  G.,  {Hainline} K.~N.,   {Hall} S.~W.,  2015, \mn@doi [\mnras]
  {10.1093/mnras/stu2341}, \href
  {http://adsabs.harvard.edu/abs/2015MNRAS.446.3492D} {446, 3492}

\bibitem[\protect\citeauthoryear{{Diamond-Stanic}, {Rieke}  \&
  {Rigby}}{{Diamond-Stanic} et~al.}{2009}]{Diamond-Stanic09}
{Diamond-Stanic} A.~M.,  {Rieke} G.~H.,   {Rigby} J.~R.,  2009, \mn@doi [\apj]
  {10.1088/0004-637X/698/1/623}, \href
  {http://adsabs.harvard.edu/abs/2009ApJ...698..623D} {698, 623}

\bibitem[\protect\citeauthoryear{{Elvis} et~al.,}{{Elvis}
  et~al.}{1994}]{elvis94}
{Elvis} M.,  et~al., 1994, \mn@doi [\apjs] {10.1086/192093}, \href
  {http://adsabs.harvard.edu/abs/1994ApJS...95....1E} {95, 1}

\bibitem[\protect\citeauthoryear{{Hainline}, {Hickox}, {Greene}, {Myers}  \&
  {Zakamska}}{{Hainline} et~al.}{2013}]{Hainline13}
{Hainline} K.~N.,  {Hickox} R.,  {Greene} J.~E.,  {Myers} A.~D.,   {Zakamska}
  N.~L.,  2013, \mn@doi [\apj] {10.1088/0004-637X/774/2/145}, \href
  {http://adsabs.harvard.edu/abs/2013ApJ...774..145H} {774, 145}

\bibitem[\protect\citeauthoryear{{Hainline}, {Hickox}, {Greene}, {Myers},
  {Zakamska}, {Liu}  \& {Liu}}{{Hainline} et~al.}{2014}]{Hainline14}
{Hainline} K.~N.,  {Hickox} R.~C.,  {Greene} J.~E.,  {Myers} A.~D.,  {Zakamska}
  N.~L.,  {Liu} G.,   {Liu} X.,  2014, \mn@doi [\apj]
  {10.1088/0004-637X/787/1/65}, \href
  {http://adsabs.harvard.edu/abs/2014ApJ...787...65H} {787, 65}

\bibitem[\protect\citeauthoryear{{Heckman} \& {Best}}{{Heckman} \&
  {Best}}{2014}]{Heckman14}
{Heckman} T.~M.,  {Best} P.~N.,  2014, \mn@doi [\araa]
  {10.1146/annurev-astro-081913-035722}, \href
  {http://adsabs.harvard.edu/abs/2014ARA%26A..52..589H} {52, 589}

\bibitem[\protect\citeauthoryear{{Heckman}, {Kauffmann}, {Brinchmann},
  {Charlot}, {Tremonti}  \& {White}}{{Heckman} et~al.}{2004}]{Heckman04}
{Heckman} T.~M.,  {Kauffmann} G.,  {Brinchmann} J.,  {Charlot} S.,  {Tremonti}
  C.,   {White} S.~D.~M.,  2004, \mn@doi [\apj] {10.1086/422872}, \href
  {http://adsabs.harvard.edu/abs/2004ApJ...613..109H} {613, 109}

\bibitem[\protect\citeauthoryear{{Isobe}, {Feigelson}, {Akritas}  \&
  {Babu}}{{Isobe} et~al.}{1990}]{Isobe90}
{Isobe} T.,  {Feigelson} E.~D.,  {Akritas} M.~G.,   {Babu} G.~J.,  1990,
  \mn@doi [\apj] {10.1086/169390}, \href
  {http://adsabs.harvard.edu/abs/1990ApJ...364..104I} {364, 104}

\bibitem[\protect\citeauthoryear{{Kauffmann} \& {Heckman}}{{Kauffmann} \&
  {Heckman}}{2009}]{Kauffmann09}
{Kauffmann} G.,  {Heckman} T.~M.,  2009, \mn@doi [\mnras]
  {10.1111/j.1365-2966.2009.14960.x}, \href
  {http://adsabs.harvard.edu/abs/2009MNRAS.397..135K} {397, 135}

\bibitem[\protect\citeauthoryear{{Kewley}, {Groves}, {Kauffmann}  \&
  {Heckman}}{{Kewley} et~al.}{2006}]{Kewley06}
{Kewley} L.~J.,  {Groves} B.,  {Kauffmann} G.,   {Heckman} T.,  2006, \mn@doi
  [\mnras] {10.1111/j.1365-2966.2006.10859.x}, \href
  {http://adsabs.harvard.edu/abs/2006MNRAS.372..961K} {372, 961}

\bibitem[\protect\citeauthoryear{{Kriss}}{{Kriss}}{1994}]{Kriss94}
{Kriss} G.,  1994, in {Crabtree} D.~R.,  {Hanisch} R.~J.,   {Barnes} J.,  eds,
  Astronomical Society of the Pacific Conference Series Vol. 61, Astronomical
  Data Analysis Software and Systems III. p.~437

\bibitem[\protect\citeauthoryear{{LaMassa}, {Heckman}, {Ptak}, {Hornschemeier},
  {Martins}, {Sonnentrucker}  \& {Tremonti}}{{LaMassa}
  et~al.}{2009}]{LaMassa09}
{LaMassa} S.~M.,  {Heckman} T.~M.,  {Ptak} A.,  {Hornschemeier} A.,  {Martins}
  L.,  {Sonnentrucker} P.,   {Tremonti} C.,  2009, \mn@doi [\apj]
  {10.1088/0004-637X/705/1/568}, \href
  {http://adsabs.harvard.edu/abs/2009ApJ...705..568L} {705, 568}

\bibitem[\protect\citeauthoryear{{LaMassa}, {Heckman}, {Ptak}, {Martins},
  {Wild}  \& {Sonnentrucker}}{{LaMassa} et~al.}{2010}]{LaMassa10}
{LaMassa} S.~M.,  {Heckman} T.~M.,  {Ptak} A.,  {Martins} L.,  {Wild} V.,
  {Sonnentrucker} P.,  2010, \mn@doi [\apj] {10.1088/0004-637X/720/1/786},
  \href {http://adsabs.harvard.edu/abs/2010ApJ...720..786L} {720, 786}

\bibitem[\protect\citeauthoryear{{Malkan}, {Gorjian}  \& {Tam}}{{Malkan}
  et~al.}{1998}]{Malkan98}
{Malkan} M.~A.,  {Gorjian} V.,   {Tam} R.,  1998, \mn@doi [\apjs]
  {10.1086/313110}, \href {http://adsabs.harvard.edu/abs/1998ApJS..117...25M}
  {117, 25}

\bibitem[\protect\citeauthoryear{{Marconi}, {Risaliti}, {Gilli}, {Hunt},
  {Maiolino}  \& {Salvati}}{{Marconi} et~al.}{2004}]{Marconi04}
{Marconi} A.,  {Risaliti} G.,  {Gilli} R.,  {Hunt} L.~K.,  {Maiolino} R.,
  {Salvati} M.,  2004, \mn@doi [\mnras] {10.1111/j.1365-2966.2004.07765.x},
  \href {http://adsabs.harvard.edu/abs/2004MNRAS.351..169M} {351, 169}

\bibitem[\protect\citeauthoryear{{Marziani}, {Sulentic}, {Dultzin-Hacyan},
  {Calvani}  \& {Moles}}{{Marziani} et~al.}{1996}]{Marziani96}
{Marziani} P.,  {Sulentic} J.~W.,  {Dultzin-Hacyan} D.,  {Calvani} M.,
  {Moles} M.,  1996, \mn@doi [\apjs] {10.1086/192291}, \href
  {http://adsabs.harvard.edu/abs/1996ApJS..104...37M} {104, 37}

\bibitem[\protect\citeauthoryear{Nemmen \& Brotherton}{Nemmen \&
  Brotherton}{2010}]{Nemmen10}
Nemmen R.~S.,  Brotherton M.~S.,  2010, \mnras, 408, 1598

\bibitem[\protect\citeauthoryear{{Nenkova}, {Sirocky}, {Ivezi{\'c}}  \&
  {Elitzur}}{{Nenkova} et~al.}{2008}]{Nenkova08}
{Nenkova} M.,  {Sirocky} M.~M.,  {Ivezi{\'c}} {\v Z}.,   {Elitzur} M.,  2008,
  \mn@doi [\apj] {10.1086/590482}, \href
  {http://adsabs.harvard.edu/abs/2008ApJ...685..147N} {685, 147}

\bibitem[\protect\citeauthoryear{{Netzer}}{{Netzer}}{2009}]{Netzer09}
{Netzer} H.,  2009, \mn@doi [\mnras] {10.1111/j.1365-2966.2009.15434.x}, \href
  {http://adsabs.harvard.edu/abs/2009MNRAS.399.1907N} {399, 1907}

\bibitem[\protect\citeauthoryear{{Netzer}, {Mainieri}, {Rosati}  \&
  {Trakhtenbrot}}{{Netzer} et~al.}{2006}]{Netzer06}
{Netzer} H.,  {Mainieri} V.,  {Rosati} P.,   {Trakhtenbrot} B.,  2006, \mn@doi
  [\aap] {10.1051/0004-6361:20054203}, \href
  {http://adsabs.harvard.edu/abs/2006A%26A...453..525N} {453, 525}

\bibitem[\protect\citeauthoryear{{Runnoe}, {Brotherton}  \& {Shang}}{{Runnoe}
  et~al.}{2012a}]{Runnoe12a}
{Runnoe} J.~C.,  {Brotherton} M.~S.,   {Shang} Z.,  2012a, \mn@doi [\mnras]
  {10.1111/j.1365-2966.2012.20620.x}, \href
  {http://adsabs.harvard.edu/abs/2012MNRAS.422..478R} {422, 478}

\bibitem[\protect\citeauthoryear{{Runnoe}, {Brotherton}  \& {Shang}}{{Runnoe}
  et~al.}{2012b}]{Runnoe12b}
{Runnoe} J.~C.,  {Brotherton} M.~S.,   {Shang} Z.,  2012b, \mn@doi [\mnras]
  {10.1111/j.1365-2966.2012.21644.x}, \href
  {http://adsabs.harvard.edu/abs/2012MNRAS.426.2677R} {426, 2677}

\bibitem[\protect\citeauthoryear{{Runnoe}, {Ganguly}, {Brotherton}  \&
  {DiPompeo}}{{Runnoe} et~al.}{2013}]{Runnoe13b}
{Runnoe} J.~C.,  {Ganguly} R.,  {Brotherton} M.~S.,   {DiPompeo} M.~A.,  2013,
  \mn@doi [\mnras] {10.1093/mnras/stt852}, \href
  {http://adsabs.harvard.edu/abs/2013MNRAS.433.1778R} {433, 1778}

\bibitem[\protect\citeauthoryear{{Schlegel}, {Finkbeiner}  \&
  {Davis}}{{Schlegel} et~al.}{1998}]{Schlegel98}
{Schlegel} D.~J.,  {Finkbeiner} D.~P.,   {Davis} M.,  1998, \mn@doi [\apj]
  {10.1086/305772}, \href {http://adsabs.harvard.edu/abs/1998ApJ...500..525S}
  {500, 525}

\bibitem[\protect\citeauthoryear{{Schwartz}}{{Schwartz}}{1978}]{Schwartz78}
{Schwartz} G.,  1978, \mn@doi [The Annals of Statistics]
  {10.1214/aos/1176344136}, 6, 461

\bibitem[\protect\citeauthoryear{{Shang} et~al.,}{{Shang}
  et~al.}{2005}]{Shang05}
{Shang} Z.,  et~al., 2005, \mn@doi [\apj] {10.1086/426134}, \href
  {http://adsabs.harvard.edu/abs/2005ApJ...619...41S} {619, 41}

\bibitem[\protect\citeauthoryear{{Shang} et~al.,}{{Shang}
  et~al.}{2011}]{Shang11}
{Shang} Z.,  et~al., 2011, \mn@doi [\apjs] {10.1088/0067-0049/196/1/2}, \href
  {http://adsabs.harvard.edu/abs/2011ApJS..196....2S} {196, 2}

\bibitem[\protect\citeauthoryear{{Shen} \& {Ho}}{{Shen} \& {Ho}}{2014}]{Shen14}
{Shen} Y.,  {Ho} L.~C.,  2014, \mn@doi [\nat] {10.1038/nature13712}, \href
  {http://adsabs.harvard.edu/abs/2014Natur.513..210S} {513, 210}

\bibitem[\protect\citeauthoryear{{Tang}, {Shang}, {Gu}, {Brotherton}  \&
  {Runnoe}}{{Tang} et~al.}{2012}]{Tang12}
{Tang} B.,  {Shang} Z.,  {Gu} Q.,  {Brotherton} M.~S.,   {Runnoe} J.~C.,  2012,
  \mn@doi [\apjs] {10.1088/0067-0049/201/2/38}, \href
  {http://adsabs.harvard.edu/abs/2012ApJS..201...38T} {201, 38}

\bibitem[\protect\citeauthoryear{{Wild} et~al.,}{{Wild} et~al.}{2011}]{Wild11}
{Wild} V.,  et~al., 2011, \mn@doi [\mnras] {10.1111/j.1365-2966.2010.17536.x},
  \href {http://adsabs.harvard.edu/abs/2011MNRAS.410.1593W} {410, 1593}

\bibitem[\protect\citeauthoryear{{Wilkes}}{{Wilkes}}{2004}]{wilkes04}
{Wilkes} B.,  2004, in {Richards} G.~T.,  {Hall} P.~B.,  eds,  Astronomical
  Society of the Pacific Conference Series Vol. 311, AGN Physics with the Sloan
  Digital Sky Survey. p.~37 (\mn@eprint {} {astro-ph/0310905})

\bibitem[\protect\citeauthoryear{{Zakamska} et~al.,}{{Zakamska}
  et~al.}{2016}]{Zakamska16}
{Zakamska} N.~L.,  et~al., 2016, \mn@doi [\mnras] {10.1093/mnras/stw718}, \href
  {http://adsabs.harvard.edu/abs/2016MNRAS.459.3144Z} {459, 3144}

\makeatother
\end{thebibliography}

\label{lastpage}
\end{document}